\documentclass[preprint2,twocolumn,times,tighten,linenumbers]{aastex631}

\usepackage{color}
\usepackage{comment}
\usepackage{enumitem}
\usepackage{amsmath,amstext,mathrsfs}

\begin{document}
\nolinenumbers
\title{Exploring magnetic fields in merging galaxy: combining polarization and velocity gradient in the Centaurus Galaxy}

\author{Quynh Lan Nguyen}
\affiliation{Department of Physics and Astronomy, University of Notre Dame, Notre Dame, IN 46556, USA}

\author[0000-0002-8455-0805]{Yue Hu*}
\affiliation{Institute for Advanced Study, 1 Einstein Drive, Princeton, NJ 08540, USA}

\author{Alex Lazarian}
\affiliation{Department of Astronomy, University of Wisconsin-Madison, Madison, WI 53706, USA}

\email{*NASA Hubble Fellow}

\begin{abstract}

\end{abstract}


\begin{abstract}
\nolinenumbers
In this study, we apply the Velocity Gradient Technique (VGT) to the merging Centaurus galaxy. We compare gradient maps derived from the PHANGS-ALMA survey using CO emission lines with magnetic field tracings from dust polarization data obtained via the HAWC+ instrument. Our analysis reveals a strong correspondence between the directions indicated by these two tracers across most of the galactic image. Specifically, we identify jet regions as areas of anti-alignment, consistent with previous reports that gradients tend to rotate 90 degrees in outflow regions. Statistically, we find that the alignment of magnetic fields, as revealed by polarization, is most accurate in regions with the highest signal-to-noise ratios. Our findings underscore the utility of velocity gradients as a valuable complementary tool for probing magnetic fields and dynamical processes in merging galaxies.

\end{abstract}


\section{Introduction} \label{sec:intro}
 
Merging plays an important role in the galaxy's evolution, so tracing the magnetic field in the merging galaxy is interesting. It can provide valuable insight into the galaxy's dynamic and evolution as well as the jet formation \citep{2021MNRAS.506..229W}. Through the interaction of the magnetic field in the merging galaxy and the magnetic field in the interstellar medium and other galaxies and cluster galaxies, we can understand better the forming and the shape of the large-scale structure of the universe \citep{2016A&A...593A..89R, 2021Natur.593...47C, 2020ApJ...892..133P}. 

The observations in far-infrared and radio for galactic and extragalactic magnetic fields have significantly developed in the past decades \citep{P-Kronberg_1994, 2015A&ARv..24....4B,2021NatAs...5..604L}. Even if it has some initial results, it still has some limitations. Dust and synchrotron polarization can trace the plane of the sky (POS) magnetic field orientation. However, dust polarization is affected by dust alignment efficiency \citep{Lazarian_2007}, and synchrotron polarization is subject to Faraday rotation \citep{2015A&ARv..24....4B}. Synchrotron polarization provides the method that can trace the magnetic fields in the warm and high temperatures of the gas \citep{ Beck:2000dc, 2011MNRAS.412.2396F, 2015A&ARv..24....4B} and dust polarization contains contribution from different gas phases, including the one associated with diffuse atomic gas and dense molecular gas \citep{Lazarian_2007, 2015ARA&A..53..501A}.

To separate the magnetic field associated with molecular gas, which is the most important fueling material for the galactic nuclei, \cite{2022ApJ...941...92H} proposed the use of the Velocity Gradient Technique (VGT; \citealt{2017ApJ...835...41G,Lazarian_2018,2018MNRAS.480.1333H}) and CO spectroscopic observations. Despite that, the study of VGT has been explored for six individual galaxies M51, NGC 1068, NGC 1097, NGC 3627, and NGC 4826 \citep{2022ApJ...941...92H}, NGC3627 \citep{2023MNRAS.519.1068L}, NGC 0628 \citep{2024ApJ...967...18Z} but its application or test to galaxy merger has not been done yet. Centaurus galaxy, as one of the closest nuclear active galaxy mergers, is the best test-bed for VGT.

In this work, we trace the magnetic field of the Centaurus galaxy by using the VGT, particularly its branch, namely, Velocity Channel Gradients (VChGs; \citealt{Lazarian_2018}) and $\rm ^{12}CO$ (J = 2–1) emission lines from the PHANGS-ALMA survey \citep{2021ApJS..257...43L}. On the other hand, the magnetic field in Centaurus has been studied by \citep{2021NatAs...5..604L} using HAWC+ dust polarization at 89$\mu$m. Here we compare the magnetic field traced by VGT to the one inferred from HAWC+.

However, it should be noted that the dynamics of galaxy mergers are more complicated than normal galaxies. This makes the interpretation of the comparison more difficult. For instance, early applications \cite{2022ApJ...941...92H,2023MNRAS.519.1068L,2024ApJ...967...18Z} of the VGT have conventionally set the turbulence injection scale, $L_{\rm inj}$, at approximately 100 pc, a measure derived from extensive turbulence studies within the Milky Way \citep{1995ApJ...443..209A, 2010ApJ...710..853C,2022ApJ...934....7H}. This assumption aligns with the typical scale at which energy is injected into the interstellar medium via stellar feedback and other dynamic processes in relatively stable galactic environments. However, in scenarios involving galaxy mergers, the dynamics of turbulence injection can be substantially altered. Mergers not only impact galactic structure and star formation but also contribute to turbulence on a much larger scale \citep{2021MNRAS.506..229W}. The merging process itself can act as a significant driver of turbulence, likely extending the injection scale well beyond the 100 pc typically observed in more quiescent galaxies like the Milky Way. Given this context, it is reasonable to hypothesize that the injection scale $L_{\rm inj}$ in the Centaurus galaxy, which has experienced dynamic merger events, could be larger than 100 pc and we expect the $\rm ^{12}CO$ observation with an average spatial resolution of $145.5$~pc satisfies the VGT's requirements for this pilot study.

The details of the archival data are shown in Section\,\ref{sec:data}. Centaurus galaxy is one of the closest nuclear active galaxies and we want to use it as a test for VGT. In Section\,\ref{sec:method}, we introduce the details theory of the VGT, velocity caustics, and VGT pipeline. 
In Section \,\ref{sec:result}, we will discuss some results about the mapping magnetic field of the Centaurus galaxy and compare it with dust polarization result. 
In Section \,\ref{sec:conclusion}, we will provide some conclusions about the magnetic fields of the Centaurus galaxy using VGT and some possibility of the galaxies merges scenarios. 

\section{Observational Data} \label{sec:data}
\subsection{Emission Lines}
In this study, we utilized the $\rm ^{12}CO$ (J = 2–1) emission lines provided by the Physics at High Angular Resolution in Nearby Galaxies (PHANGS)–ALMA survey \citep{2021ApJS..257...43L}. The PHANGS-ALMA survey offers $\rm ^{12}CO$ (J = 1 - 0) emission line data at a spatial resolution of approximately $8.13''$ (about 145.5 pc) and a velocity resolution of 2.5 km s$^{-1}$ for the nearby galaxy Centaurus. This survey achieves a high signal-to-noise ratio, with an RMS brightness temperature noise level around $\sim0.30\pm0.13$ K km s$^{-1}$.

\subsection{Polarization measurement}
For polarization measurements, we employed data from the High-resolution Airborne Wideband Camera Plus (HAWC+) archived in the HAWC+ database \citep{2018JAI.....740008H,2021NatAs...5..604L}. Magnetic field orientations were determined using the relation $\phi_B=\phi+\pi/2$, where $\phi$ represents the polarization angle. Measurements were taken from band C (89 $\mu$m, FWHM $\approx7.8''$).

We limited our analysis to pixels where the polarization fraction ($p$) divided by its uncertainty ($\sigma_p$) exceeds 1 ($p/\sigma_p > 1$). To compare with the magnetic field measured with VGT, we smooth the polarization data to FWHM $\approx40''$ matching the spatial resolution of VGT.

\section{Methodology} \label{sec:method}

\subsection{Theoretical consideration: Velocity Gradient Technique}
The theoretical underpinning of the Velocity Gradient Technique (VGT) lies in the anisotropy inherent in MHD turbulence, a concept extensively developed by \citet{GS95} and \citet{LV99}. This anisotropy suggests that turbulent eddies elongate along magnetic field lines, manifesting notably in local reference frames. \citet{LV99} formalized this anisotropy as follows:
\begin{equation}
\label{eq.lv99}
l_\parallel = L_{\rm inj}\left(\frac{l_\bot}{L_{\rm inj}}\right)^{\frac{2}{3}}M_A^{-4/3},~~~M_A\le 1,
\end{equation}
where $l_\bot$ and $l_\parallel$ are the scales perpendicular and parallel to the local magnetic field, respectively. Here, $M_A = v_{\rm inj}/v_A$ represents the Alfv\'en Mach number, with $v_{\rm inj}$ being the injection velocity and $v_A$ the Alfv\'en speed.

This anisotropy has been demonstrated by numerical simulations \citep{CV20, 2001ApJ...554.1175M, CL03, 2010ApJ...720..742K, HXL21} and in situ measurements in the solar wind \citep{2016ApJ...816...15W, 2020FrASS...7...83M, 2021ApJ...915L...8D,2024ApJ...962...89Z}. From Eq.\ref{eq.lv99}, we understand that $l_\parallel$ significantly exceeds $l_\bot$, leading to the dominance of the gradient of velocity fluctuations perpendicular to the local magnetic field:
\begin{equation}
\label{eq.vg}
\nabla v_l \approx \frac{v_{l,\bot}}{l_{\bot}} = \frac{v_{\rm inj}}{L_{\rm inj}}M_A^{1/3}\left(\frac{l_{\bot}}{L_{\rm inj}}\right)^{-2/3}, M_A \le 1.
\end{equation}

This scaling relation ($\propto l_\bot^{-2/3}$) is crucial as it reveals that the amplitude of the turbulent velocity gradient increases at smaller scales, which is a characteristic not typically observed in non-turbulent velocity fields such as galactic differential rotation. For effective application to external galaxies, observations should resolve the scale at which turbulence's anisotropy manifests prominently, typically at $L_{\rm inj}$ for sub-Alfv\'en and trans-Alfv\'en turbulence or $L_{\rm inj}M_A^{-3}$ for super-Alfv\'enic turbulence \citep{2006ApJ...645L..25L}. 
\begin{figure*}[htbp!]
\centering
\includegraphics[width=0.8\linewidth]{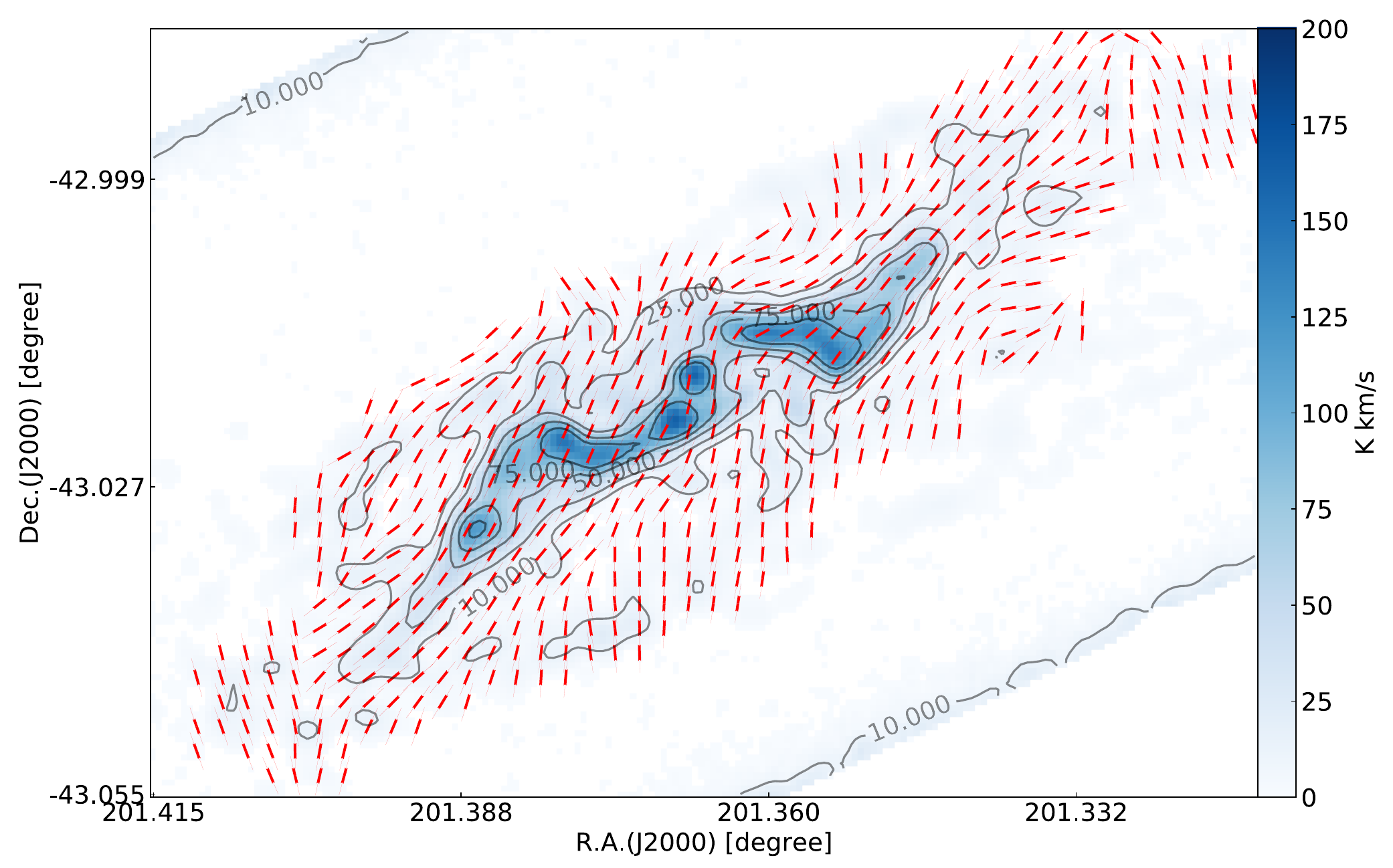}
\caption{The magnetic field of Centaurus mapped with VGT using the $\rm CO$ (2-1) emission line. The magnetic field segments are laid upon Centaurus' CO emission intensity map.
\label{fig:vgt}}
\end{figure*}

\begin{figure*}[htbp!]
\centering
\includegraphics[width=0.8\linewidth]{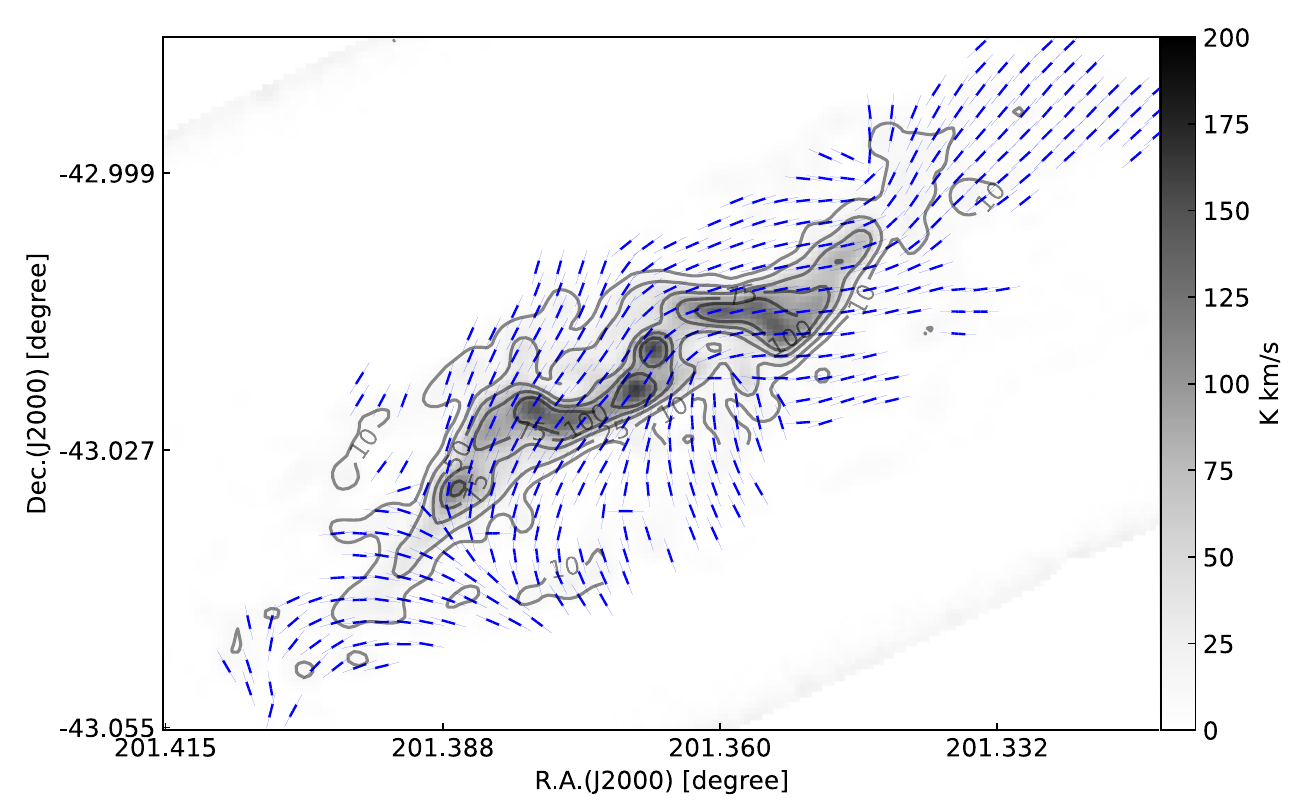}
\caption{Same as Fig.~\ref{fig:vgt}, but the magnetic field is measured by HAWC+ polarization at $89~\mu$m.
\label{fig:hawc+}}
\end{figure*}

\subsection{Velocity caustics}
The anisotropic nature of MHD turbulence can be observed in velocity channel maps, $p(x, y, v_{\rm los})$, obtained from spectroscopic observations. This observation is primarily due to the velocity caustics effect, which posits that real density structures with different velocities may be sampled into the same velocity channel, thereby significantly altering the observed intensity structures within the spectroscopic channel maps \citep{LP00,LP04,2016MNRAS.461.1227K,2023MNRAS.524.2994H}.

Mathematically, the intensity observed in a channel can be expressed as the sum of two terms \citep{2023MNRAS.524.2994H}:
\begin{equation}
p(x,y,v_{\rm los})=p_{v}(x,y,v_{\rm los})+p_{d}(x,y,v_{\rm los}),    
\end{equation}
where $p_{v}$ represents fluctuations primarily caused by velocity—known as the velocity caustics effect—while $p_{dc}$ reflects the inhomogeneities within the actual density field. The relative contribution of these terms is dependent on the channel width, i.e., the velocity resolution of the observation. A narrower channel width enhances the contribution of $p_{v}$, leading to a scenario where intensity fluctuations in thin channels are predominantly due to velocity fluctuations. Specifically, when the channel width $\Delta v$ is narrower than the velocity dispersion of the turbulent eddies under observation, the intensity fluctuations within such a thin channel will be primarily driven by velocity fluctuations.

Turbulence studies within the Milky Way suggest that the velocity dispersion at a scale of 100 pc is approximately 10 km/s \citep{1995ApJ...443..209A, 2010ApJ...710..853C,2022ApJ...934....7H}. This dispersion can be even higher in environments such as galaxy mergers. Given that the PHANGS-ALMA survey offers a velocity resolution of 2.5 km/s, this ensures that the velocity caustics dominate the observed intensity in the spectroscopic velocity channels.

\subsection{VGT pipeline}
In this study, we employed the VGT, following a multi-step process detailed in \citep{2022ApJ...941...92H,2023MNRAS.519.1068L}:\\
\textbf{ Step 1.} Each thin channel map, $p(x, y, v_{\rm los})$, undergoes convolution with 3$\times$3 Sobel kernels $G_x$ and $G_y$ to compute the gradients:
\begin{equation}
\begin{aligned}
\nabla_{x} p(x, y, v_{\rm los}) &= G_x * p(x, y, v_{\rm los}), \\
\nabla_{y} p(x, y, v_{\rm los}) &= G_y * p(x, y, v_{\rm los}),
\end{aligned}
\end{equation}
where * denotes convolution. These gradients are used to determine the pixelized gradient map $\psi_g (x, y, v_{\rm los})$:
\begin{equation}
\psi_g (x, y, v_{\rm los}) = \tan^{-1}\left(\frac{\nabla_{y} p(x, y, v_{\rm los})}{\nabla_{x}p(x, y, v_{\rm los})}\right).
\end{equation}
Only pixels with a brightness temperature less than three times the RMS noise are retained for subsequent analysis.

\textbf{Step 2.} The pixelized map $\psi_g(x, y, v_{\rm los})$ is segmented into 20$\times$20 pixel sub-blocks—although other sizes have been tested \citep{YL17a,LY18a,2020ApJ...905..129H} —and processed as follows:
\begin{enumerate}
\item A histogram of gradient orientations within each sub-block is produced and fitted with a Gaussian distribution.
\item The peak of this Gaussian distribution is taken as the dominant gradient orientation for that sub-block.
\end{enumerate}

\textbf{Step 3.} Following sub-block averaging, the averaged gradient angle map $\psi_{gs}(x, y, v_{\rm los})$ is compiled for each velocity channel. Repeating the Steps 1 and 2 for every channel within the velocity range of interest, we can then construct the pseudo-Stokes parameters $Q_g(x,y)$ and $U_g(x,y)$:
\begin{equation}
\label{eq.QU}
\begin{aligned}
Q_{\rm g} (x,y) &= \int_{v_{\rm los, min}}^{v_{\rm los, max}} p(x, y, v_{\rm los}) \cos(2\psi_{gs}(x,y,v_{\rm los})) dv_{\rm los},\\
U_{\rm g} (x,y) &= \int_{v_{\rm los, min}}^{v_{\rm los, max}} p(x, y, v_{\rm los}) \sin(2\psi_{gs}(x,y,v_{\rm los})) dv_{\rm los},
\end{aligned}
\end{equation}
where $v_{\rm los, max}$ and $v_{\rm los, min}$ represent the upper and lower levels of the velocity range used for integration. The POS magnetic field orientation is inferred from:
\begin{equation}
\psi_{\rm B}(x,y) = \frac{1}{2} \tan^{-1}\left(\frac{U_{\rm g} (x,y)}{Q_{\rm g} (x,y)}\right) + \frac{\pi}{2}.
\end{equation}

To quantify the agreement between VGT and the magnetic field inferred from polarization, we utilize the Alignment Measure (AM; \citealt{GL17}), expressed as:
\begin{equation}
\begin{aligned}
{\rm AM} = 2 \left(\cos^2\theta_{\rm r} - \frac{1}{2}\right)
\end{aligned}
\end{equation}
Here, $\theta_{\rm r} = \vert\phi_{\rm B}-\psi_{\rm B}\vert$, where $\phi_{\rm B}$ is the magnetic field angle inferred from polarization. An AM value of 1 implies parallel alignment of $\phi_{\rm B}$ and $\psi_{\rm B}$, while -1 indicates perpendicularity.

\begin{figure*}[htbp!]
\centering
\includegraphics[width=0.99\linewidth]{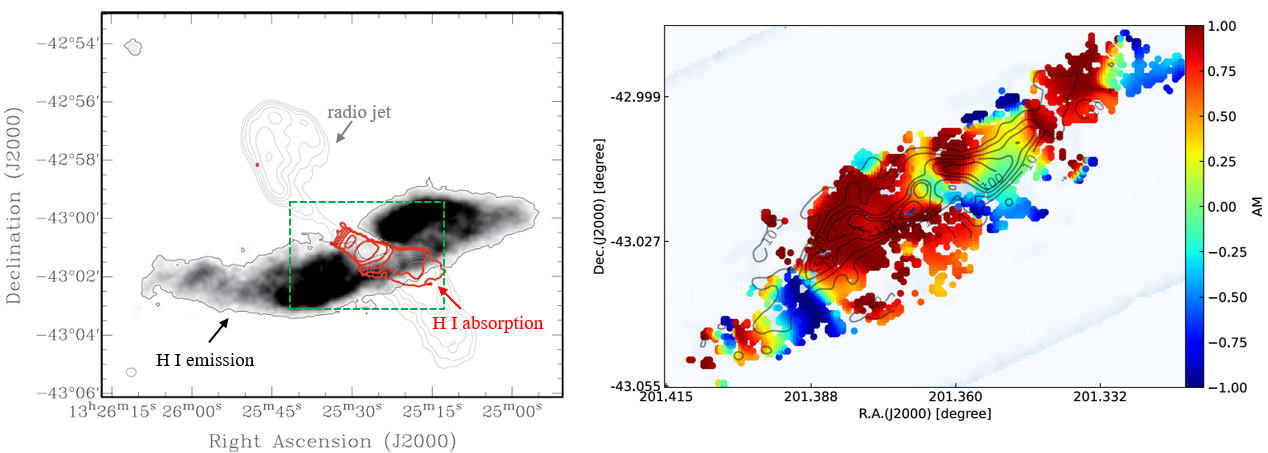}
\caption{Left: H I emission in greyscale overlaid with radio continuum contours (grey) and H I absorption contours (red). The green box highlights the region where we are studying the VGT-CO. Reproduced with permission \copyright ESO, \cite{2010A&A...515A..67S}, A\&A, 515, A67. Right: The spatial distributions of the AM between VGT-CO measurement and HAWC+ polarization with a resolution of FWHM $40''$. 
\label{fig:am}}
\end{figure*}

\section{Result}  
\label{sec:result}

\begin{figure}[htbp!]
\centering
\includegraphics[width=1.0\linewidth]{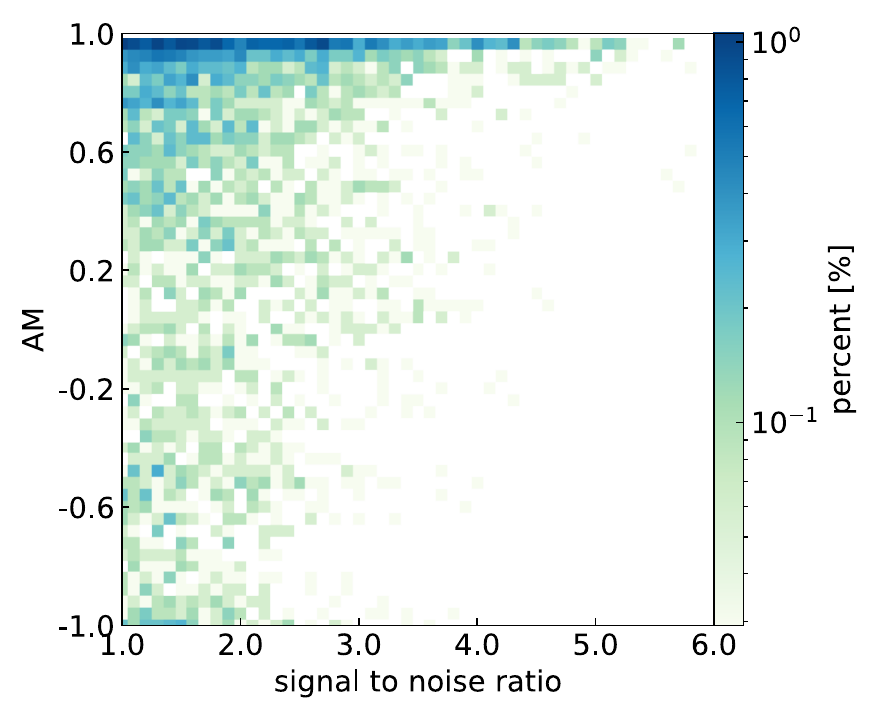}
\caption{ A 2D histogram of the dust polarization's signal-to-noise ratio and AM between VGT and polarization
\label{fig:am_hist}}
\end{figure}
Fig.~\ref{fig:vgt} presents the magnetic fields of Centaurus as mapped by VGT-CO. The image captures the twisting of the magnetic field from the southeast and northwest filamentary structures, which resemble spiral-arm-like features, towards the molecular disk and into the nucleus. The magnetic field, as depicted by VGT-CO, primarily aligns with these two filamentary structures. However, it exhibits a twist across the central region of approximately 500 pc radius around the nucleus. Additionally, the magnetic fields inferred from HAWC+ polarization, illustrated in Fig.~\ref{fig:hawc+}, exhibit similar morphology and generally agree with the measurements from VGT-CO. Notably, near the southeast tail, the polarization-inferred magnetic field diverges from the galactic structure indicated by CO, instead crossing it. This specific pattern is absent in the VGT-CO measurements. 

The misalignment between the magnetic fields traced by VGT-CO and those indicated by dust polarization is quantified by the AM, as shown in Fig.~\ref{fig:am}. The anti-alignment, indicated by negative AM values, is observed in the two tails, while a misalignment with AM $\approx 0$ is noted in the northwest of the central disk. A higher degree of the mean alignment is observed in several Seyfert galaxies and the Milky Way \citep{2022ApJ...941...92H,2023MNRAS.519.1068L,2024ApJ...967...18Z}. We attribute this effect to the fact that the Centaurus is a merging galaxy with much more complex dynamics of gas. Jets, shocks, and inflows affect the gradients, especially if the measurements. These effects were discussed at length in \cite{Lazarian_2018,2019ApJ...886...17H,2020ApJ...897..123H,2024MNRAS.530.1066L}. In particular, anti-alignment is present in the radio jet regions, which is expected for the outflow regions. Further studies of the origin of this discrepancy are required. This can elucidate the dynamics of gas in the merging galaxy. 

Another source of the observed misalignment can be the difference in CO and dust distribution in these regions. Unlike CO, dust is well mixed with all interstellar phases, including molecular and atomic gas \citep{2015ARA&A..53..501A}. In Fig.~\ref{fig:am}, we see the central low-AM region is associated with strong HI absorption, suggesting the dust polarization in that region might be dominated by dust associated with H I rather than CO. In addition, HAWC+ dust polarization at $89~\mu$m is anticipated to trace magnetic fields in warmer media than CO. If magnetic fields in warm and cold gas are very different, there is a warmer phase preferentially traced by HAWC+; the morphology of magnetic fields with low-AM could reflect that the magnetic fields in warm and cold gas are different. 

We also note that the most significant misalignment between VGT-CO and dust polarization measurements is in the low-density regions associated with poor signal-to-noise ratio. This is illustrated in Fig.~\ref{fig:am_hist}, where we observe that we do not observe a negative AM at the high signal-to-noise ratio regions.  

Overall, the global agreement between the magnetic field orientations measured with VGT-CO and those inferred from dust polarization, especially in the central disk, underscores the coherence of magnetic fields across different gas phases. This includes the cold molecular phase traced by CO and warmer phases indicated by HAWC+. The observed coherence across these phases suggests that the magnetic fields threading the multiphase gas, spanning a range of densities, are integrally involved in the galactic dynamic evolution. This supports the notion that molecular clouds are part of a unified magnetic ecosystem in galaxies and that magnetic fields in multiphase gas maintain an overall coherent structure even in a merging galaxy. Additional input from synchrotron polarization could be useful for getting more information. In the aforementioned studies of Seyfert galaxies with VGT  \citep{2022ApJ...941...92H,2023MNRAS.519.1068L,2024ApJ...967...18Z}, the alignment between the velocity gradients in CO and synchrotron could be better than the alignment with dust, which elucidated the 3D magnetic field structure.

\section{Conclusion}  \label{sec:conclusion}
Velocity gradients have demonstrated efficacy in tracing magnetic fields within magnetohydrodynamic (MHD) turbulence environments. This capability has been validated across various regions of the Milky Way and several Seyfert galaxies. In this paper, we extend the application of velocity gradient techniques (VGT) to the more complex scenario of a merging galaxy. Specifically, we measure velocity gradients in the Centaurus galaxy and evaluate their effectiveness in tracing magnetic fields, using dust polarization as a comparative measure. Our findings indicate that VGT results generally concur with dust polarization on a statistical basis. Additionally, we identify localized regions of significant misalignment and anti-alignment. Drawing on prior studies, we hypothesize that outflows and the presence of shocked gas may account for these discrepancies. If corroborated, this comparison of gradients and polarization could elucidate the underlying physics of galactic mergers. Furthermore, our study aids in pinpointing regions within disturbed galaxies where velocity gradients can reliably map magnetic fields in the absence of other tracers. Future research, utilizing higher signal-to-noise data and incorporating synchrotron polarization, is essential to further explore these findings.

\begin{acknowledgments}
A.L. acknowledges the support of NASA ATP AAH7546, NSF grants AST 2307840, and ALMA SOSPADA-016. Financial support for this work was provided by NASA through award 09\_0231 issued by the Universities Space Research Association, Inc. (USRA). Y.H. acknowledges the support for this work provided by NASA through the NASA Hubble Fellowship grant No. HST-HF2-51557.001 awarded by the Space Telescope Science Institute, which is operated by the Association of Universities for Research in Astronomy, Incorporated, under NASA contract NAS5-26555. This work used SDSC Expanse CPU at SDSC through allocations PHY230032, PHY230033, PHY230091, and PHY230105 from the Advanced Cyberinfrastructure Coordination Ecosystem: Services \& Support (ACCESS) program, which is supported by National Science Foundation grants \#2138259, \#2138286, \#2138307, \#2137603, and \#2138296. Q.L.N. acknowledges the support in part by grant NSF PHY-1748958 to the Kavli Institute for Theoretical Physics (KITP), the Institute for  Cosmic  Ray  Research-International University Research Program, Notre Dame College of Science Internal Grant. 
\end{acknowledgments}

%

\vspace{5mm}


\software{Julia \citep{bezanson2012julia} 
          }






\bibliography{sample631}{}

\begin{thebibliography}{}
\expandafter\ifx\csname natexlab\endcsname\relax\def\natexlab#1{#1}\fi
\providecommand{\url}[1]{\href{#1}{#1}}
\providecommand{\dodoi}[1]{doi:~\href{http://doi.org/#1}{\nolinkurl{#1}}}
\providecommand{\doeprint}[1]{\href{http://ascl.net/#1}{\nolinkurl{http://ascl.net/#1}}}
\providecommand{\doarXiv}[1]{\href{https://arxiv.org/abs/#1}{\nolinkurl{https://arxiv.org/abs/#1}}}

\bibitem[{{Andersson} {et~al.}(2015){Andersson}, {Lazarian}, \&
  {Vaillancourt}}]{2015ARA&A..53..501A}
{Andersson}, B.~G., {Lazarian}, A., \& {Vaillancourt}, J.~E. 2015, \araa, 53,
  501, \dodoi{10.1146/annurev-astro-082214-122414}

\bibitem[{{Armstrong} {et~al.}(1995){Armstrong}, {Rickett}, \&
  {Spangler}}]{1995ApJ...443..209A}
{Armstrong}, J.~W., {Rickett}, B.~J., \& {Spangler}, S.~R. 1995, \apj, 443,
  209, \dodoi{10.1086/175515}

\bibitem[{Beck(2001)}]{Beck:2000dc}
Beck, R. 2001, Space Sci. Rev., 99, 243, \dodoi{10.1023/A:1013805401252}

\bibitem[{{Beck}(2015)}]{2015A&ARv..24....4B}
{Beck}, R. 2015, \aapr, 24, 4, \dodoi{10.1007/s00159-015-0084-4}

\bibitem[{Bezanson {et~al.}(2012)Bezanson, Karpinski, Shah, \&
  Edelman}]{bezanson2012julia}
Bezanson, J., Karpinski, S., Shah, V.~B., \& Edelman, A. 2012, arXiv preprint
  arXiv:1209.5145

\bibitem[{{Chepurnov} \& {Lazarian}(2010)}]{2010ApJ...710..853C}
{Chepurnov}, A., \& {Lazarian}, A. 2010, \apj, 710, 853,
  \dodoi{10.1088/0004-637X/710/1/853}

\bibitem[{{Chibueze} {et~al.}(2021){Chibueze}, {Sakemi}, {Ohmura}, {Machida},
  {Akamatsu}, {Akahori}, {Nakanishi}, {Parekh}, {van Rooyen}, \&
  {Takeuchi}}]{2021Natur.593...47C}
{Chibueze}, J.~O., {Sakemi}, H., {Ohmura}, T., {et~al.} 2021, \nat, 593, 47,
  \dodoi{10.1038/s41586-021-03434-1}

\bibitem[{{Cho} \& {Lazarian}(2003)}]{CL03}
{Cho}, J., \& {Lazarian}, A. 2003, \mnras, 345, 325,
  \dodoi{10.1046/j.1365-8711.2003.06941.x}

\bibitem[{{Cho} \& {Vishniac}(2000)}]{CV20}
{Cho}, J., \& {Vishniac}, E.~T. 2000, \apj, 539, 273, \dodoi{10.1086/309213}

\bibitem[{{Duan} {et~al.}(2021){Duan}, {He}, {Bowen}, {Woodham}, {Wang},
  {Chen}, {Mallet}, \& {Bale}}]{2021ApJ...915L...8D}
{Duan}, D., {He}, J., {Bowen}, T.~A., {et~al.} 2021, ApJL, 915, L8

\bibitem[{{Fletcher} {et~al.}(2011){Fletcher}, {Beck}, {Shukurov},
  {Berkhuijsen}, \& {Horellou}}]{2011MNRAS.412.2396F}
{Fletcher}, A., {Beck}, R., {Shukurov}, A., {Berkhuijsen}, E.~M., \&
  {Horellou}, C. 2011, \mnras, 412, 2396,
  \dodoi{10.1111/j.1365-2966.2010.18065.x}

\bibitem[{{Goldreich} \& {Sridhar}(1995)}]{GS95}
{Goldreich}, P., \& {Sridhar}, S. 1995, \apj, 438, 763, \dodoi{10.1086/175121}

\bibitem[{{Gonz{\'a}lez-Casanova} \&
  {Lazarian}(2017{\natexlab{a}})}]{2017ApJ...835...41G}
{Gonz{\'a}lez-Casanova}, D.~F., \& {Lazarian}, A. 2017{\natexlab{a}}, \apj,
  835, 41, \dodoi{10.3847/1538-4357/835/1/41}

\bibitem[{{Gonz{\'a}lez-Casanova} \& {Lazarian}(2017{\natexlab{b}})}]{GL17}
---. 2017{\natexlab{b}}, \apj, 835, 41

\bibitem[{{Ha} {et~al.}(2022){Ha}, {Li}, {Kounkel}, {Xu}, {Li}, \&
  {Zheng}}]{2022ApJ...934....7H}
{Ha}, T., {Li}, Y., {Kounkel}, M., {et~al.} 2022, \apj, 934, 7,
  \dodoi{10.3847/1538-4357/ac76bf}

\bibitem[{{Harper} {et~al.}(2018){Harper}, {Runyan}, {Dowell}, {Wirth},
  {Amato}, {Ames}, {Amiri}, {Banks}, {Bartels}, {Benford}, {Berthoud},
  {Buchanan}, {Casey}, {Chapman}, {Chuss}, {Cook}, {Derro}, {Dotson}, {Evans},
  {Fixsen}, {Gatley}, {Guerra}, {Halpern}, {Hamilton}, {Hamlin}, {Hansen},
  {Heimsath}, {Hermida}, {Hilton}, {Hirsch}, {Hollister}, {Hostetter}, {Irwin},
  {Jhabvala}, {Jhabvala}, {Kastner}, {Kov{\'a}cs}, {Lin}, {Loewenstein},
  {Looney}, {Lopez-Rodriguez}, {Maher}, {Michail}, {Miller}, {Moseley},
  {Novak}, {Pernic}, {Rennick}, {Rhody}, {Sandberg}, {Sandford}, {Santos},
  {Shafer}, {Sharp}, {Shirron}, {Siah}, {Silverberg}, {Sparr}, {Spotz},
  {Staguhn}, {Toorian}, {Towey}, {Tuttle}, {Vaillancourt}, {Voellmer},
  {Volpert}, {Wang}, \& {Wollack}}]{2018JAI.....740008H}
{Harper}, D.~A., {Runyan}, M.~C., {Dowell}, C.~D., {et~al.} 2018, Journal of
  Astronomical Instrumentation, 7, 1840008

\bibitem[{{Hu} {et~al.}(2023){Hu}, {Lazarian}, {Alina}, {Pogosyan}, \&
  {Ho}}]{2023MNRAS.524.2994H}
{Hu}, Y., {Lazarian}, A., {Alina}, D., {Pogosyan}, D., \& {Ho}, K.~W. 2023,
  \mnras, 524, 2994, \dodoi{10.1093/mnras/stad1924}

\bibitem[{{Hu} {et~al.}(2022){Hu}, {Lazarian}, {Beck}, \&
  {Xu}}]{2022ApJ...941...92H}
{Hu}, Y., {Lazarian}, A., {Beck}, R., \& {Xu}, S. 2022, \apj, 941, 92,
  \dodoi{10.3847/1538-4357/ac9df0}

\bibitem[{{Hu} {et~al.}(2020{\natexlab{a}}){Hu}, {Lazarian}, \&
  {Bialy}}]{2020ApJ...905..129H}
{Hu}, Y., {Lazarian}, A., \& {Bialy}, S. 2020{\natexlab{a}}, \apj, 905, 129,
  \dodoi{10.3847/1538-4357/abc3c6}

\bibitem[{{Hu} {et~al.}(2020{\natexlab{b}}){Hu}, {Lazarian}, \&
  {Yuen}}]{2020ApJ...897..123H}
{Hu}, Y., {Lazarian}, A., \& {Yuen}, K.~H. 2020{\natexlab{b}}, \apj, 897, 123,
  \dodoi{10.3847/1538-4357/ab9948}

\bibitem[{{Hu} {et~al.}(2021){Hu}, {Xu}, \& {Lazarian}}]{HXL21}
{Hu}, Y., {Xu}, S., \& {Lazarian}, A. 2021, \apj, 911, 37,
  \dodoi{10.3847/1538-4357/abea18}

\bibitem[{{Hu} {et~al.}(2018){Hu}, {Yuen}, \& {Lazarian}}]{2018MNRAS.480.1333H}
{Hu}, Y., {Yuen}, K.~H., \& {Lazarian}, A. 2018, \mnras, 480, 1333,
  \dodoi{10.1093/mnras/sty1807}

\bibitem[{{Hu} {et~al.}(2019){Hu}, {Yuen}, \& {Lazarian}}]{2019ApJ...886...17H}
---. 2019, \apj, 886, 17, \dodoi{10.3847/1538-4357/ab4b5e}

\bibitem[{{Kandel} {et~al.}(2016){Kandel}, {Lazarian}, \&
  {Pogosyan}}]{2016MNRAS.461.1227K}
{Kandel}, D., {Lazarian}, A., \& {Pogosyan}, D. 2016, \mnras, 461, 1227,
  \dodoi{10.1093/mnras/stw1296}

\bibitem[{{Kowal} \& {Lazarian}(2010)}]{2010ApJ...720..742K}
{Kowal}, G., \& {Lazarian}, A. 2010, \apj, 720, 742,
  \dodoi{10.1088/0004-637X/720/1/742}

\bibitem[{Kronberg(1994)}]{P-Kronberg_1994}
Kronberg, P.~P. 1994, Reports on Progress in Physics, 57, 325,
  \dodoi{10.1088/0034-4885/57/4/001}

\bibitem[{{Lazarian}(2006)}]{2006ApJ...645L..25L}
{Lazarian}, A. 2006, \apjl, 645, L25, \dodoi{10.1086/505796}

\bibitem[{Lazarian \& Hoang(2007)}]{Lazarian_2007}
Lazarian, A., \& Hoang, T. 2007, The Astrophysical Journal, 669, L77,
  \dodoi{10.1086/523849}

\bibitem[{{Lazarian} \& {Pogosyan}(2000)}]{LP00}
{Lazarian}, A., \& {Pogosyan}, D. 2000, \apj, 537, 720, \dodoi{10.1086/309040}

\bibitem[{{Lazarian} \& {Pogosyan}(2004)}]{LP04}
---. 2004, \apj, 616, 943, \dodoi{10.1086/422462}

\bibitem[{{Lazarian} \& {Vishniac}(1999)}]{LV99}
{Lazarian}, A., \& {Vishniac}, E.~T. 1999, \apj, 517, 700,
  \dodoi{10.1086/307233}

\bibitem[{Lazarian \& Yuen(2018)}]{Lazarian_2018}
Lazarian, A., \& Yuen, K.~H. 2018, The Astrophysical Journal, 853, 96,
  \dodoi{10.3847/1538-4357/aaa241}

\bibitem[{{Lazarian} \& {Yuen}(2018)}]{LY18a}
{Lazarian}, A., \& {Yuen}, K.~H. 2018, \apj, 853, 96,
  \dodoi{10.3847/1538-4357/aaa241}

\bibitem[{{Leroy} {et~al.}(2021){Leroy}, {Schinnerer}, {Hughes}, {Rosolowsky},
  {Pety}, {Schruba}, {Usero}, {Blanc}, {Chevance}, {Emsellem}, {Faesi},
  {Herrera}, {Liu}, {Meidt}, {Querejeta}, {Saito}, {Sandstrom}, {Sun},
  {Williams}, {Anand}, {Barnes}, {Behrens}, {Belfiore}, {Benincasa},
  {Be{\v{s}}li{\'c}}, {Bigiel}, {Bolatto}, {den Brok}, {Cao}, {Chandar},
  {Chastenet}, {Chiang}, {Congiu}, {Dale}, {Deger}, {Eibensteiner}, {Egorov},
  {Garc{\'\i}a-Rodr{\'\i}guez}, {Glover}, {Grasha}, {Henshaw}, {Ho}, {Kepley},
  {Kim}, {Klessen}, {Kreckel}, {Koch}, {Kruijssen}, {Larson}, {Lee}, {Lopez},
  {Machado}, {Mayker}, {McElroy}, {Murphy}, {Ostriker}, {Pan}, {Pessa},
  {Puschnig}, {Razza}, {S{\'a}nchez-Bl{\'a}zquez}, {Santoro}, {Sardone},
  {Scheuermann}, {Sliwa}, {Sormani}, {Stuber}, {Thilker}, {Turner}, {Utomo},
  {Watkins}, \& {Whitmore}}]{2021ApJS..257...43L}
{Leroy}, A.~K., {Schinnerer}, E., {Hughes}, A., {et~al.} 2021, \apjs, 257, 43,
  \dodoi{10.3847/1538-4365/ac17f3}

\bibitem[{{Liu} {et~al.}(2024){Liu}, {Hu}, \& {Lazarian}}]{2024MNRAS.530.1066L}
{Liu}, M., {Hu}, Y., \& {Lazarian}, A. 2024, \mnras, 530, 1066,
  \dodoi{10.1093/mnras/stae863}

\bibitem[{{Liu} {et~al.}(2023){Liu}, {Hu}, {Lazarian}, {Xu}, \&
  {Soida}}]{2023MNRAS.519.1068L}
{Liu}, M., {Hu}, Y., {Lazarian}, A., {Xu}, S., \& {Soida}, M. 2023, \mnras,
  519, 1068, \dodoi{10.1093/mnras/stac3518}

\bibitem[{{Lopez-Rodriguez}(2021)}]{2021NatAs...5..604L}
{Lopez-Rodriguez}, E. 2021, Nature Astronomy, 5, 604,
  \dodoi{10.1038/s41550-021-01329-9}

\bibitem[{{Maron} \& {Goldreich}(2001)}]{2001ApJ...554.1175M}
{Maron}, J., \& {Goldreich}, P. 2001, \apj, 554, 1175, \dodoi{10.1086/321413}

\bibitem[{{Matteini} {et~al.}(2020){Matteini}, {Franci}, {Alexandrova},
  {Lacombe}, {Landi}, {Hellinger}, {Papini}, \&
  {Verdini}}]{2020FrASS...7...83M}
{Matteini}, L., {Franci}, L., {Alexandrova}, O., {et~al.} 2020, Frontiers in
  Astronomy and Space Sciences, 7, 83

\bibitem[{{Paliya} {et~al.}(2020){Paliya}, {P{\'e}rez}, {Garc{\'\i}a-Benito},
  {Ajello}, {Prada}, {Alberdi}, {Suh}, {Chandra}, {Dom{\'\i}nguez}, {Marchesi},
  {Di Matteo}, {Hartmann}, \& {Chiaberge}}]{2020ApJ...892..133P}
{Paliya}, V.~S., {P{\'e}rez}, E., {Garc{\'\i}a-Benito}, R., {et~al.} 2020,
  \apj, 892, 133, \dodoi{10.3847/1538-4357/ab754f}

\bibitem[{{Rodenbeck} \& {Schleicher}(2016)}]{2016A&A...593A..89R}
{Rodenbeck}, K., \& {Schleicher}, D. R.~G. 2016, \aap, 593, A89,
  \dodoi{10.1051/0004-6361/201527393}

\bibitem[{{Struve} {et~al.}(2010){Struve}, {Oosterloo}, {Morganti}, \&
  {Saripalli}}]{2010A&A...515A..67S}
{Struve}, C., {Oosterloo}, T.~A., {Morganti}, R., \& {Saripalli}, L. 2010,
  \aap, 515, A67, \dodoi{10.1051/0004-6361/201014355}

\bibitem[{{Wang} {et~al.}(2016){Wang}, {Tu}, {Marsch}, {He}, \&
  {Wang}}]{2016ApJ...816...15W}
{Wang}, X., {Tu}, C., {Marsch}, E., {He}, J., \& {Wang}, L. 2016, \apj, 816,
  15, \dodoi{10.3847/0004-637X/816/1/15}

\bibitem[{{Whittingham} {et~al.}(2021){Whittingham}, {Sparre}, {Pfrommer}, \&
  {Pakmor}}]{2021MNRAS.506..229W}
{Whittingham}, J., {Sparre}, M., {Pfrommer}, C., \& {Pakmor}, R. 2021, \mnras,
  506, 229, \dodoi{10.1093/mnras/stab1425}

\bibitem[{{Yuen} \& {Lazarian}(2017)}]{YL17a}
{Yuen}, K.~H., \& {Lazarian}, A. 2017, \apjl, 837, L24,
  \dodoi{10.3847/2041-8213/aa6255}

\bibitem[{{Zhao} {et~al.}(2024{\natexlab{a}}){Zhao}, {Zhou}, {Baan}, {Hu},
  {Lazarian}, {Tang}, {Esimbek}, {He}, {Li}, {Ji}, {Chang}, \&
  {Tursun}}]{2024ApJ...967...18Z}
{Zhao}, M., {Zhou}, J., {Baan}, W.~A., {et~al.} 2024{\natexlab{a}}, \apj, 967,
  18, \dodoi{10.3847/1538-4357/ad3a62}

\bibitem[{{Zhao} {et~al.}(2024{\natexlab{b}}){Zhao}, {Yan}, {Liu}, {Yuen}, \&
  {Shi}}]{2024ApJ...962...89Z}
{Zhao}, S., {Yan}, H., {Liu}, T.~Z., {Yuen}, K.~H., \& {Shi}, M.
  2024{\natexlab{b}}, \apj, 962, 89, \dodoi{10.3847/1538-4357/ad132e}

\end{thebibliography}
\bibliographystyle{aasjournal}


\end{document}